\documentclass[prd,twocolumn,showpacs,preprintnumbers,amsmath,amssymb]{revtex4}
\usepackage{graphicx}
\begin{document}
\preprint{INT-PUB-10-013}

\title{Fisher's zeros as boundary of renormalization group  flows in complex coupling spaces}
\author{A. Denbleyker}
\author{Daping Du}
\author{Yuzhi Liu}
\author{Y. Meurice}
%\email{yannick-meurice@uiowa.edu}
\author{Haiyuan Zou}
\affiliation{Department of Physics and Astronomy\\ The University of Iowa\\
Iowa City, Iowa 52242, USA }
\date{\today}
\def\tc{\lambda^t}
 \def\mn{\mathcal{N}_p}

\begin{abstract}
We propose new methods to extend the renormalization group transformation to complex coupling spaces. We argue that the Fisher's  zeros
are located at the boundary of the complex basin of attraction of infra-red fixed points.  We support this picture with numerical calculations at finite volume for two-dimensional $O(N)$ models in the large-$N$ limit and the hierarchical Ising model.
We present numerical evidence that, as the volume increases,  the Fisher's zeros of 4-dimensional pure gauge $SU(2)$ lattice gauge theory with a Wilson action, stabilize at a distance larger than 0.15 from the real axis in the complex $\beta=4/g^2$ plane. We discuss the implications for proofs of confinement and searches for nontrivial infra-red fixed points in models beyond the standard model.
\end{abstract}
\maketitle
\def\gf{\mathfrak{B} }

The renormalization group (RG) method has played a prominent role in  advancing our understanding of strongly interacting and strongly correlated systems. A question of central importance in this context is 
to find out if the physical spectrum of gauge theories, sigma models or 
Hubbard models contains massless particles, such as gauge or Nambu-Goldstone bosons, or if a mass gap is generated dynamically. For asymptotically  free theories, this question can be  addressed by studying the (marginally relevant) RG flows coming out of the weakly coupled fixed point. 
If these RG flows end at the strongly coupled fixed point, the correlations among local observables decay exponentially with the separation 
(mass gap) and Wilson loops decay exponentially with their area (confinement). Showing rigorously that this statement correct is 
a possible strategy \cite{Tomboulis:2009zz}
to prove confinement in non-abelian gauge theories on the lattice. 

Confinement can be lost by either introducing a finite temperature or enough matter fields to modify the running of the coupling constant. In particular,  if enough species of fermions are added without spoiling asymptotic freedom, 
one may expect a nontrivial infrared (IR) fixed point with conformal symmetry \cite{banks81}.  Recently, there has been a renewed interest in this possibility to build models for possible new physics 
beyond the standard model.  Various 
extensions of QCD  have been studied \cite{shamir08}.  Other scenarios involving multiple confinement/deconfinement transitions 
were also proposed in Refs. \cite{Myers:2007vc}. Establishing the existence of an IR fixed point is often controversial and it would be desirable to find criteria independent of the RG method used.  

In this Letter, we show that considerable insight on these questions can be obtained by extending the RG flows in 
complex coupling space.  We provide empirical evidence  that the global behavior of the complex RG flows can be determined  by simply 
calculating the complex zeros of the partition function in the complex coupling plane (Fisher's zeros) and related singular points. In the large volume limit, we argue that the basin  of attraction of the strongly coupled (confining) fixed point is delimited by Fisher's zeros. The complex conjugated pair of zeros closest to the real axis can be seen 
as a ``gate'' controlling the complex flows between the weakly and  strongly coupled fixed points. Monitoring the position of these zeros as the volume increases can provide a way to decide if the theory is confining or not. For $SU(2)$ pure gauge theory this is a challenging task \cite{quasig} especially when we increase the volume. In the following, we report the first numerically stable calculations of Fisher 
zeros on a $6^4$ lattice. Details will be provided elsewhere \cite{su2progress}.

In the study of flows or differential equations, it is often  enlightening to consider their complexification. Recent RG studies \cite{kaplan09} discuss the loss of conformal invariance and disappearance of fixed points in the complex plane when 
a parameter is varied beyond some critical value. In addition, analytic continuation of a theory into the complex coupling plane is an essential tool to understand the large-order behavior of perturbative series. It has been used to explain why perturbative series have a zero 
radius of convergence \cite{dyson52} and to determine accurately the 
growth of the perturbative series \cite{bender69,leguillou90}. 
For lattice models with compact integration \cite{gluodyn04,Meurice:2009bq}, there is a change but not a loss of vacuum when the real positive coupling $g_0^2$ changes sign. This may explain the apparent power growth behavior of weak coupling series obtained with 
stochastic perturbation theory \cite{direnzo2000,third,npp}. A  complex RG perspective on these questions would be very desirable. 

In the following, we consider  three types of models known for their absence of phase transition: the non-linear $O(N)$ sigma model on a square lattice in the large-$N$ limit, the two-dimensional Ising hierarchical model and $SU(2)$ lattice gauge theory in 4 dimensions with a Wilson action. 
We use generic notations which can be used interchangeably for the 3 models. $\beta$ denotes the inverse 't Hooft coupling $1/(g_0^2N)$  for the $O(N)$ model, the inverse temperature for the hierarchical model, and $4/g_0^2$ for the $SU(2)$ gauge theory. These three models are discussed in more detail in 
Refs. \cite{Meurice:2009bq,hmreview,Denbleyker:2008ss} respectively. We use the notation $a$ for the lattice spacing, $m_G$ for the mass gap and $M\equiv am_G$ its dimensionless  form.
$\beta$ should not be confused with the Callan-Symanzik $\beta$ function which will be denoted $\beta_{CS}$. 
In general, $M^2\partial \beta/\partial M^2\propto \beta_{CS}/g_0^3$, with a 
model dependent positive constant of proportionality. 
For asymptotically free theories, in the small $a$ limit we have 
\begin{equation}
\label{eq:af} 
\beta(M^2)\simeq A+B\ {\rm ln}(1/M^2)
\end{equation}  with $B>0$, proportional to minus the first coefficient of the $\beta_{CS}$-function. 
\def\mn{\mathcal{N}}We will also use the generic spectral decomposition \cite{alves89,Denbleyker:2008ss} of the partition function:
\begin{equation}
Z =\int_0^{S_{max}}dSn(S){\rm e}^{-\beta S}\ ,
\label{eq:intds}
\end{equation}
where $S$ is the total action (or energy).  We call $n(S)$ the density of states. We use the notation ${\rm ln}(n(S))/\mn\equiv f(S/\mn)$ with 
$\mn$ the number of lattice sites for spin models and the number of plaquettes for gauge models.
 
Simple one-dimensional RG flows can be generated  
when the $\beta$ dependence on the lattice spacing $a$ at fixed physical mass gap $m_G$ is known. Beyond the celebrated asymptotic scaling regime of Eq. (\ref{eq:af}), local polynomial parametrizations 
of ${\rm ln}(M^2)$ are known in $SU(3)$ lattice gauge theory
\cite{guagnelli98}, but their analytical continuation is only valid in a small region. For the $O(N)$ models at infinite volume and infinite $N$, we have a close  form expression \cite{PhysRev.86.821}
valid for any real positive $M^2$:
\begin{equation}
\label{eq:gap}
\beta(M^2)=
\int_{-\pi}^{\pi}\int_{-\pi}^{\pi}\frac{d^2k}{(2\pi)^2}\frac{1}{P^2(k_1,k_2)+M^2\ ,}
\end{equation}
with $P^2(k_1,k_2)\equiv2(2-{\rm cos}(k_1)-{\rm cos}(k_2))$. 
In the limit of small $M^2$, we have the asymptotic form of Eq. (\ref{eq:af}) with $B=1/(4\pi)$. 

\begin{figure}
\includegraphics[width=3.in,angle=0]{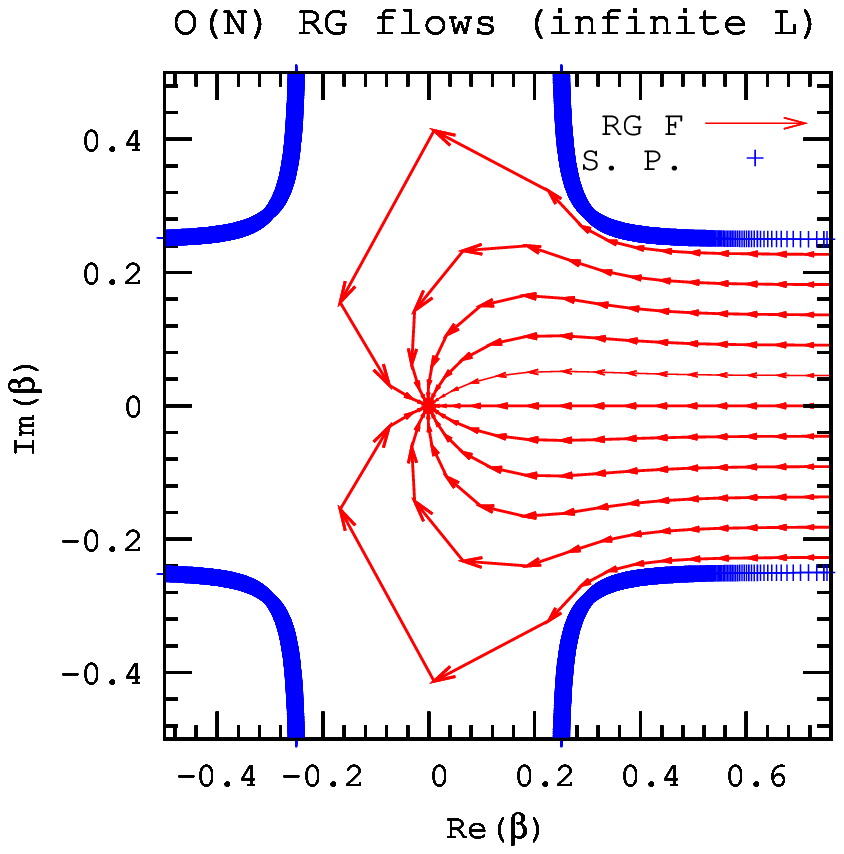}
\includegraphics[width=3.in,angle=0]{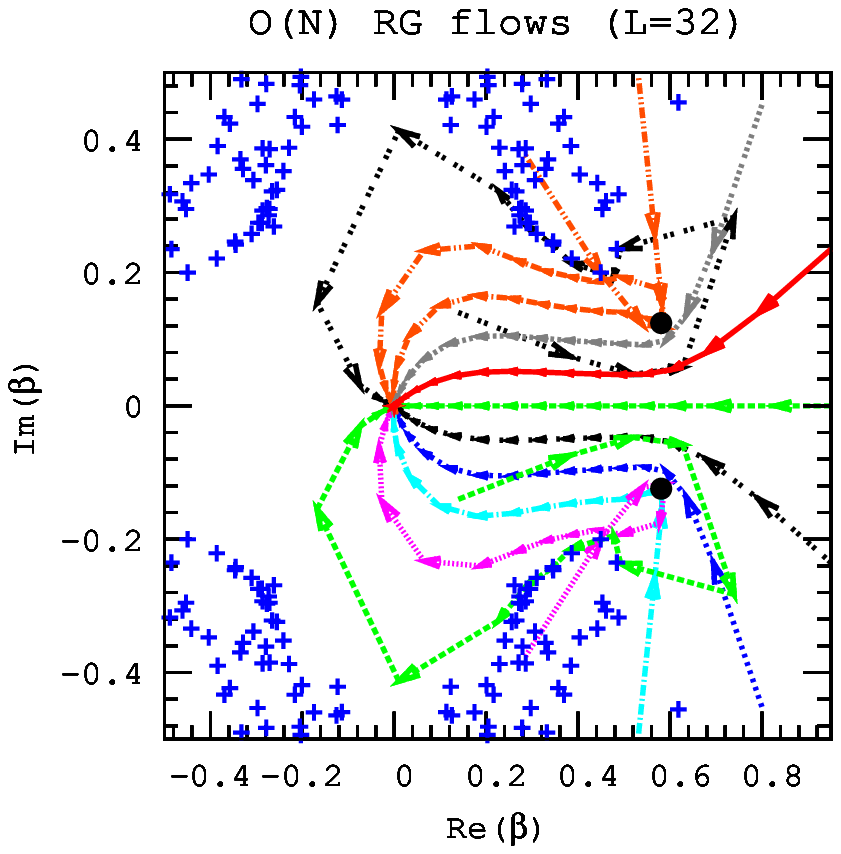}
\caption{\label{fig:inf} Top: Infinite $L$ RG flows (arrows). The blending small crosses (blue on-line) are the $\beta$ images of two lines of points located very close above and below the $[-8,0]$ cut. Bottom: same procedure and initial conditions but for $L=32$; the crosses are the images of the singular points.  The image of the CSP described in the text appear as two large filled circles.}
\end{figure}

The mapping can be analytically continued to the cut complex $M^2$ plane with a cut running from -8 to 0. The image of this cut plane 
in the complex $\beta$ plane is an asymptotically cross-shaped region partially shown in Fig. \ref{fig:inf}. In Ref. \cite{Meurice:2009bq} it was argued that the Fisher's zeros should lay outside of the this crossed shaped region. 
Complex RG flows can be obtained by increasing the lattice spacing (kept real) with a fixed complex $m_G$. Fig. \ref{fig:inf} shows the RG flows for 11 initial values of $M^2$ taken on a small circle around the origin and then multiplied repeatedly by a factor 2. 
The flows stay inside the image of the $[-8,0]$ cut. By taking initial values very close to  the real negative axis, it is possible to follow closely this boundary. 

Lattice simulations are performed at finite volume and it is important to 
understand the modification of the simple flow picture provided above when the volume is finite.  For  $L\times L$  lattices with periodic boundary conditions, the integral in Eq. (\ref{eq:gap}) is replaced 
by an average over the $L^2$ momenta coming in units of $2\pi/L$. The mapping $\beta(M^2)$ becomes a rational function and its inversion requires a Riemann surface with a number of sheets scaling like $L^2$.
The cuts (in the $\beta$ plane) end at the images $\beta(M^2_{sing.})$ of singular points $M^2_{sing.}$ such that 
$\beta'(M^2_{sing.})=0$. 
These values of $\beta$ can be interpreted as the  complex zeros of the non-perturbative $\beta_{CS}$-function \cite{Meurice:2009bq}. 
If we consider the image of a straight line coming out of the origin in the complex $M^2$ plane and making an angle $\theta$ with the positive real axis, its image may wrap around a certain number of the $\beta(M^2_{sing.})$.  A detailed analysis \cite{onprogress} shows 
that it only occurs when $|\theta|>\pi/2$. The complex conjugate images with the smallest $\theta$  play an essential role in the understanding of the global properties of the flows.
In practice, their $\theta$ is very close to $\pm \pi/2$ and we call them 
the closest singular points (CSP).  The situation is illustrated in the bottom part of  Fig. \ref{fig:inf} for $L=32$. 
The procedure is identical to the infinite volume case, but some results are strinkingly different. 
For very small $|M^2|$, the pole at 0 in the finite volume sum that replaces the integral in Eq. (\ref{eq:gap}), dominates and replaces the logarithmic divergence at infinite volume. Consequently, the image of 
a small circle around the origin in the $M^2$ plane is a large circle in the $\beta$ plane. 
There are 
288 singular points. 
%The image of the CSP is represented by two . 
The flows corresponding to $|\theta| <\pi/2$ go between 
the images of the CSP (large filled circles) from the right. The flows corresponding to $|\theta| >\pi/2$ go between 
the images of the CSP from the left, wrap around the  images of the CSP and eventually end up 
at zero. In summary, the global properties of the RG flows are controlled by the CSP. In the large volume limit,  the real part of the image of the CSP goes to infinity and the imaginary part stabilizes at 
$(\pi/2)(1/(4\pi))=1/8$ as can be inferred from Eq. (\ref{eq:af}). This stabilization implies that the complex RG flows on the positive real axis and the neighboring complex flows can reach the strongly coupled fixed point without obstruction. 

\begin{figure}
\includegraphics[width=3.in,angle=0]{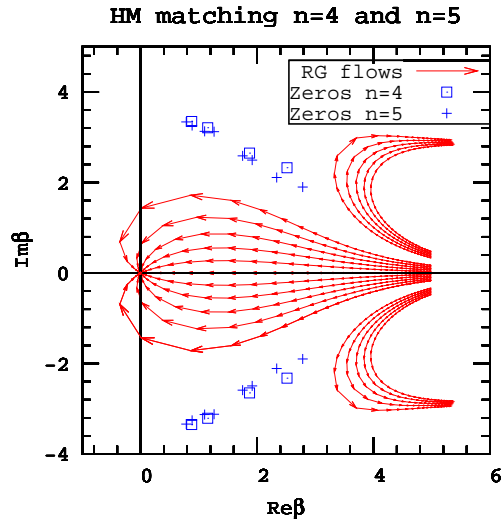}
\caption{\label{fig:hm}Unambiguous RG flows for the hierarchical model in the complex $\beta$ plane  obtained by the two lattice method. The crosses and open boxes are at the Fishers zeros for $2^4$ and $2^5$ sites. }
\end{figure}

Complex RG flows can also be calculated by extending  2-lattice matching methods to the case of complex $\beta$. In the following, we use a slightly modified version of Ref. \cite{PhysRevB.27.1736,Hasenfratz:1984hx}. The idea is to consider large distance observables that can be calculated on lattices with different sizes in lattice spacing units but 
equal physical sizes.  The large distance behavior is probed by calculating the correlations between two large neighbor blocks $B$ and $NB$ of physical size $|B|$. The ratio of the block volume to the total physical volume $V$ is the same for both lattices.  
In order to bypass the determination of the field rescaling, we consider the ratio
\begin{equation}
R(\beta,V/a^D)\equiv\frac{\left\langle (\sum_{x\in B}\vec{\phi}_x )(\sum_{y\in NB}\vec{\phi}_y)\right\rangle_\beta}{\left\langle (\sum_{x\in B}\vec{\phi}_x)(\sum_{y\in B}\vec{\phi}_y ))\right\rangle_\beta} \ .
\end{equation}
In Ref.  \cite{Hasenfratz:1984hx}, the whole ratio was averaged. Here,  we use blocked observables that depend linearly on the original variables, the average at complex $\beta$ can be defined by reweighting configurations at real $\beta$. 
A discrete RG transformation mapping $\beta$ into $\beta'$ while the lattice spacing changes from $a$ to $ba$  is obtained by requiring the matching 
$R(\beta,V/a^D)=R(\beta',V/(ba)^D)$.
When $\beta$ is complex, there are typically many $\beta'$. In special 
cases, the matching condition reduces to polynomial equations in which the multivaluedness can be addressed systematically. 
For practical purposes, one would like to be able to use Newton's method to construct the RG flows. This works if there is only one $\beta'$ solution close to $\beta$. For the two spin models considered here, we found out that unless the RG flow get near the Fisher's zeros, 
the distance $|\beta - \beta'|$ singles out one $\beta'$ unambiguously. 
The situation is illustrated for the hierarchical model in Fig. \ref{fig:hm}. 
We required  $R(\beta,2^5)=R(\beta',2^4)$ using the exactly calculable probability distribution for blocks covering half the volume. 29 initial $\beta$ were chosen on a line with constant $Re\beta=5$. The 8 trajectories 
passing by the Fisher's zeros led to ambiguous choices of $\beta'$ and are not displayed. More detail on this method will be provided in Refs. \cite{hmprogress,onprogress}.

\begin{figure}
\includegraphics[width=3.in]{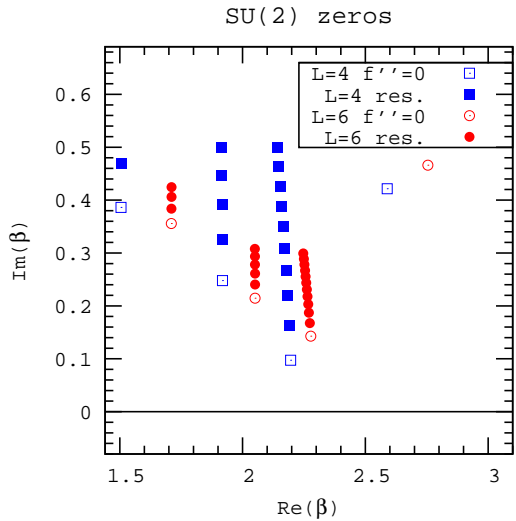}
\caption{\label{fig:zeros}  Images of the zeros of $f''(s)$ in the $\beta$ plane (open symbols) and Fisher's zeros (filled symbols) for $4^4$ (squares) and $6^4$ (circles) lattices.}
\end{figure}

We now introduce a generic method to restrict the region where Fisher's zeros can be found. Using the density of states, it is clear that the contributions at fixed $\beta$  should come from a small region near the saddle point. 
\begin{eqnarray}
n(S){\rm e}^{-\beta\mn s}&=&{\rm e}^{\mn (f(s)-\beta s)}  \\ &=&{\rm e}^{\mn (f(s_0)+(1/2)f''(s_0)(s-s_0)^2+\dots)} \nonumber \  ,
\end{eqnarray}with $s=S/\mn$ and $f'(s_0)=\beta$. 
As long as $Ref''(s_0)<0$, the distribution becomes Gaussian in the infinite volume because if we define $\mn f''(s_0)(s-s_0)^2\equiv y^2$ as the normal variable, higher order in $s-s_0$ in the exponential will be suppressed by negative powers of $\mn$. Gaussian distributions have no complex zeros \cite{alves90b}, and consequently, we could look for the level curve $Ref''(s_0)=0$ as the boundary of the region where Fisher's zeros may appear. 
In Fig. 1 of Ref. \cite{Bazavov:2009wz}  the regions where $Ref''(s_0)\geq0$ are depicted as  narrow ``tongues'' coming vertically toward the real axis.  
In the $U(1)$ case, a conjugate pair pinches the real axis, 
but for $SU(2)$ a finite gap remains present.  This suggests that the Fisher's zeros of these models should appear on approximately vertical linear structures. For $SU(2)$, the imaginary part of Fisher's are too large to use simple reweighting methods \cite{quasig}. By using Chebyshev interpolation for $f(s)$ and monitoring the numerical stability of the integrals with the residue theorem \cite{Meurice:2009bq}, it is possible to obtain reasonably stable results \cite{su2progress} that confirm this picture (see  Fig. \ref{fig:zeros}). Unlike the $U(1)$ case,  the imaginary part of the lowest zeros does not decrease as the volume increases, but their linear density increases at a rate compatible with $L^{-4}$.

For $O(N)$ models, it is possible to write close form expressions of the partition function in the approximation (justified in the large-$N$ limit) where we only keep the zero mode of the auxiliary field enforcing the constraint $\vec{\phi}.\vec{\phi}=1$.
For $L$ and $N$ not too large, it is possible to use the residue theorem to calculate exactly the integral. From this exact expression, one can calculate the Fisher's zero and the density of states which happens to be piecewise polynomial. Using the exact form of the density of state, we can calculate the zeros of $f''$.  As in the case of the $SU(2)$ gauge theory, approximately vertical lines of zeros appear above the singular points of the two mappings discussed above. Their linear density increases at a rate compatible with 
$L^{-4}$  \cite{onprogress}.

In summary, we have shown with examples that RG flows 
in one real coupling can be extended to the complex coupling plane. 
As the volume increases, the stabilization of the Fisher's zeros away from the real axis allows the complex flows to reach the strongly coupled fixed point. Two-lattice matching methods can be 
extended to the complex plane by reweighting existing gauge or spin configurations (as long as the imaginary part of $\beta$ is not too large). We plan to apply this method to decide if extensions of QCD confine or not. 

Part of this work was done during the workshop 
``New applications of the renormalization group
method in nuclear, particle and condensed matter physics" held at the Institute for Nuclear Theory, University of Washington, Seattle. 
We thank the participants of this workshop and also  A. Bazavov, C. Bender and A. Velytsky 
for stimulating discussions. 
This 
research was supported in part  by the Department of Energy
under Contract No. FG02-91ER40664.
%\end{acknowledgments}
%\bibliography{macbib}

\begin{thebibliography}{39}
\expandafter\ifx\csname natexlab\endcsname\relax\def\natexlab#1{#1}\fi
\expandafter\ifx\csname bibnamefont\endcsname\relax
  \def\bibnamefont#1{#1}\fi
\expandafter\ifx\csname bibfnamefont\endcsname\relax
  \def\bibfnamefont#1{#1}\fi
\expandafter\ifx\csname citenamefont\endcsname\relax
  \def\citenamefont#1{#1}\fi
\expandafter\ifx\csname url\endcsname\relax
  \def\url#1{\texttt{#1}}\fi
\expandafter\ifx\csname urlprefix\endcsname\relax\def\urlprefix{URL }\fi
\providecommand{\bibinfo}[2]{#2}
\providecommand{\eprint}[2][]{\url{#2}}

  \bibitem[{\citenamefont{Tomboulis}(2009)}]{Tomboulis:2009zz}
\bibinfo{author}{\bibfnamefont{E.~T.} \bibnamefont{Tomboulis}},
  \bibinfo{journal}{Mod. Phys. Lett.} \textbf{\bibinfo{volume}{A24}},
  \bibinfo{pages}{2717} (\bibinfo{year}{2009}).


\bibitem[{\citenamefont{Banks and Zaks}(1982)}]{banks81}
\bibinfo{author}{\bibfnamefont{T.}~\bibnamefont{Banks}} \bibnamefont{and}
  \bibinfo{author}{\bibfnamefont{A.}~\bibnamefont{Zaks}},
  \bibinfo{journal}{Nucl. Phys.} \textbf{\bibinfo{volume}{B196}},
  \bibinfo{pages}{189} (\bibinfo{year}{1982}).

\bibitem[{\citenamefont{Shamir et~al.}(2008)\citenamefont{Shamir, Svetitsky,
  and DeGrand}}]{shamir08}
\bibinfo{author}{\bibfnamefont{Y.}~\bibnamefont{Shamir}},
  \bibinfo{author}{\bibfnamefont{B.}~\bibnamefont{Svetitsky}},
  \bibnamefont{and} \bibinfo{author}{\bibfnamefont{T.}~\bibnamefont{DeGrand}},
  \bibinfo{journal}{Phys. Rev.} \textbf{\bibinfo{volume}{D78}},
  \bibinfo{pages}{031502} (\bibinfo{year}{2008}); %\eprint{0803.1707}.
%\bibitem[{\citenamefont{Appelquist et~al.}(2009)\citenamefont{Appelquist,
%  Fleming, and Neil}}]{appelquist09}
\bibinfo{author}{\bibfnamefont{T.}~\bibnamefont{Appelquist}},
  \bibinfo{author}{\bibfnamefont{G.~T.} \bibnamefont{Fleming}},
  \bibnamefont{and} \bibinfo{author}{\bibfnamefont{E.~T.} \bibnamefont{Neil}},
  \bibinfo{journal}{Phys. Rev.} \textbf{\bibinfo{volume}{D79}},
  \bibinfo{pages}{076010} (\bibinfo{year}{2009});  %\eprint{0901.3766}.
%
%\bibitem[{\citenamefont{Hasenfratz}(2009)}]{hasenfratz09}
\bibinfo{author}{\bibfnamefont{A.}~\bibnamefont{Hasenfratz}},
  \bibinfo{journal}{Phys. Rev.} \textbf{\bibinfo{volume}{D80}},
  \bibinfo{pages}{034505} (\bibinfo{year}{2009}); %, \eprint{0907.0919}.
%
%\bibitem[{\citenamefont{Fodor et~al.}(2009)\citenamefont{Fodor, Holland, Kuti,
%  Nogradi, and Schroeder}}]{fodor09}
\bibinfo{author}{\bibfnamefont{Z.}~\bibnamefont{Fodor}},
  \bibinfo{author}{\bibfnamefont{K.}~\bibnamefont{Holland}},
  \bibinfo{author}{\bibfnamefont{J.}~\bibnamefont{Kuti}},
  \bibinfo{author}{\bibfnamefont{D.}~\bibnamefont{Nogradi}}, \bibnamefont{and}
  \bibinfo{author}{\bibfnamefont{C.}~\bibnamefont{Schroeder}},
  \bibinfo{journal}{Phys. Lett.} \textbf{\bibinfo{volume}{B681}},
  \bibinfo{pages}{353} (\bibinfo{year}{2009}); %, \eprint{0907.4562}.
%
%$\bibitem[{\citenamefont{Deuzeman et~al.}(2009)\citenamefont{Deuzeman, Lombardo,
 % and Pallante}}]{deuzeman09}
\bibinfo{author}{\bibfnamefont{A.}~\bibnamefont{Deuzeman}},
  \bibinfo{author}{\bibfnamefont{M.~P.} \bibnamefont{Lombardo}},
  \bibnamefont{and} \bibinfo{author}{\bibfnamefont{E.}~\bibnamefont{Pallante}}
  (\bibinfo{year}{2009}). %, \eprint{0904.4662}.

\bibitem[{\citenamefont{Myers and Ogilvie}(2008)}]{Myers:2007vc}
\bibinfo{author}{\bibfnamefont{J.~C.} \bibnamefont{Myers}} \bibnamefont{and}
  \bibinfo{author}{\bibfnamefont{M.~C.} \bibnamefont{Ogilvie}},
  \bibinfo{journal}{Phys. Rev.} \textbf{\bibinfo{volume}{D77}},
  \bibinfo{pages}{125030} (\bibinfo{year}{2008}); \bibinfo{author}{\bibfnamefont{H.}~\bibnamefont{Nishimura}} \bibnamefont{and}
  \bibinfo{author}{\bibfnamefont{M.~C.} \bibnamefont{Ogilvie}},
  \bibinfo{journal}{Phys. Rev.} \textbf{\bibinfo{volume}{D81}},
  \bibinfo{pages}{014018} (\bibinfo{year}{2010}).%, \eprint{0911.2696}.
  
   \bibitem[{\citenamefont{Denbleyker et~al.}(2007)\citenamefont{Denbleyker, Du,
  Meurice, and Velytsky}}]{quasig}
\bibinfo{author}{\bibfnamefont{A.}~\bibnamefont{Denbleyker}},
  \bibinfo{author}{\bibfnamefont{D.}~\bibnamefont{Du}},
  \bibinfo{author}{\bibfnamefont{Y.}~\bibnamefont{Meurice}}, \bibnamefont{and}
  \bibinfo{author}{\bibfnamefont{A.}~\bibnamefont{Velytsky}},
  \bibinfo{journal}{Phys. Rev.} \textbf{\bibinfo{volume}{D76}},
  \bibinfo{pages}{116002} (\bibinfo{year}{2007}).% \eprint{arXiv:0708.0438[hep-lat]}.


\bibitem[{\citenamefont{Denbleyker et~al.}()\citenamefont{Denbleyker, Daping,
  Meurice, and Velytsky}}]{su2progress}
\bibinfo{author}{\bibfnamefont{A.}~\bibnamefont{Denbleyker}},
  \bibinfo{author}{\bibfnamefont{D.}~\bibnamefont{Du}},
  \bibinfo{author}{\bibfnamefont{Y.}~\bibnamefont{Meurice}}, \bibnamefont{and}
  \bibinfo{author}{\bibfnamefont{A.}~\bibnamefont{Velytsky}},
  \emph{\bibinfo{title}{Fisher's zeros of SU(2) lattice gauge theory}},
  \bibinfo{note}{preprint in preparation}.

\bibitem[{\citenamefont{Kaplan et~al.}(2009)\citenamefont{Kaplan, Lee, Son, and
  Stephanov}}]{kaplan09}
 \bibinfo{author}{\bibfnamefont{D.~B.} \bibnamefont{Kaplan}},
  \bibinfo{author}{\bibfnamefont{J.-W.} \bibnamefont{Lee}},
  \bibinfo{author}{\bibfnamefont{D.~T.} \bibnamefont{Son}}, \bibnamefont{and}
  \bibinfo{author}{\bibfnamefont{M.~A.} \bibnamefont{Stephanov}},
  \bibinfo{journal}{Phys. Rev.} \textbf{\bibinfo{volume}{D80}},
  \bibinfo{pages}{125005} (\bibinfo{year}{2009}); \bibinfo{author}{\bibfnamefont{S.}~\bibnamefont{Moroz}} \bibnamefont{and}
  \bibinfo{author}{\bibfnamefont{R.}~\bibnamefont{Schmidt}},
  \bibinfo{journal}{Annals Phys.} \textbf{\bibinfo{volume}{325}},
  \bibinfo{pages}{491} (\bibinfo{year}{2010}).
\bibitem[{\citenamefont{Dyson}(1952)}]{dyson52}
\bibinfo{author}{\bibfnamefont{F.}~\bibnamefont{Dyson}},
  \bibinfo{journal}{Phys. Rev.} \textbf{\bibinfo{volume}{85}},
  \bibinfo{pages}{631} (\bibinfo{year}{1952}).

\bibitem[{\citenamefont{Bender and Wu}(1969)}]{bender69}
\bibinfo{author}{\bibfnamefont{C.}~\bibnamefont{Bender}} \bibnamefont{and}
  \bibinfo{author}{\bibfnamefont{T.~T.} \bibnamefont{Wu}},
  \bibinfo{journal}{Phys. Rev.} \textbf{\bibinfo{volume}{184}},
  \bibinfo{pages}{1231} (\bibinfo{year}{1969}).

\bibitem[{\citenamefont{LeGuillou and Zinn-Justin}(1990)}]{leguillou90}
\bibinfo{author}{\bibfnamefont{J.~C.} \bibnamefont{LeGuillou}}
  \bibnamefont{and}
  \bibinfo{author}{\bibfnamefont{J.}~\bibnamefont{Zinn-Justin}},
  \emph{\bibinfo{title}{Large-Order Behavior of Perturbation Theory}}
  (\bibinfo{publisher}{North Holland}, \bibinfo{address}{Amsterdam},
  \bibinfo{year}{1990}).

\bibitem[{\citenamefont{Li and Meurice}(2005)}]{gluodyn04}
\bibinfo{author}{\bibfnamefont{L.}~\bibnamefont{Li}} \bibnamefont{and}
  \bibinfo{author}{\bibfnamefont{Y.}~\bibnamefont{Meurice}},
  \bibinfo{journal}{Phys. Rev. D} \textbf{\bibinfo{volume}{71}},
  \bibinfo{pages}{016008} (\bibinfo{year}{2005}), \eprint{hep-lat/0410029}.

\bibitem[{\citenamefont{Meurice}(2009)}]{Meurice:2009bq}
\bibinfo{author}{\bibfnamefont{Y.}~\bibnamefont{Meurice}},
  \bibinfo{journal}{Phys. Rev.} \textbf{\bibinfo{volume}{D80}},
  \bibinfo{pages}{054020} (\bibinfo{year}{2009}).

\bibitem[{\citenamefont{Di~Renzo and Scorzato}(2001)}]{direnzo2000}
\bibinfo{author}{\bibfnamefont{F.}~\bibnamefont{Di~Renzo}} \bibnamefont{and}
  \bibinfo{author}{\bibfnamefont{L.}~\bibnamefont{Scorzato}},
  \bibinfo{journal}{JHEP} \textbf{\bibinfo{volume}{10}}, \bibinfo{pages}{038}
  (\bibinfo{year}{2001}); 
\bibinfo{author}{\bibfnamefont{P.~E.~L.} \bibnamefont{Rakow}},
  \bibinfo{journal}{PoS} \textbf{\bibinfo{volume}{LAT2005}},
  \bibinfo{pages}{284} (\bibinfo{year}{2006}); \bibinfo{author}{\bibfnamefont{E.~M.} \bibnamefont{Ilgenfritz}}
  \bibnamefont{et~al.}  \bibinfo{journal}{PoS} \textbf{\bibinfo{volume}{LAT2009}},
  \bibinfo{pages}{236}(\bibinfo{year}{2009}).
  
\bibitem[{\citenamefont{Li and Meurice}(2006)}]{third}
\bibinfo{author}{\bibfnamefont{L.}~\bibnamefont{Li}} \bibnamefont{and}
  \bibinfo{author}{\bibfnamefont{Y.}~\bibnamefont{Meurice}},
  \bibinfo{journal}{Phys. Rev.} \textbf{\bibinfo{volume}{D73}},
  \bibinfo{pages}{036006} (\bibinfo{year}{2006}).%, \eprint{hep-lat/0507034}.

\bibitem[{\citenamefont{Meurice}(2006)}]{npp}
\bibinfo{author}{\bibfnamefont{Y.}~\bibnamefont{Meurice}},
  \bibinfo{journal}{Phys. Rev.} \textbf{\bibinfo{volume}{D74}},
  \bibinfo{pages}{096005} (\bibinfo{year}{2006}).%, \eprint{hep-lat/0609005}.
  
  \bibitem[{\citenamefont{Meurice}(2007)}]{hmreview}
\bibinfo{author}{\bibfnamefont{Y.}~\bibnamefont{Meurice}}, \bibinfo{journal}{J.
  Phys.} \textbf{\bibinfo{volume}{A40}}, \bibinfo{pages}{R39}
  (\bibinfo{year}{2007}).%, \eprint{hep-th/0701191}.



\bibitem[{\citenamefont{Denbleyker et~al.}(2008)\citenamefont{Denbleyker, Du,
  Liu, Meurice, and Velytsky}}]{Denbleyker:2008ss}
\bibinfo{author}{\bibfnamefont{A.}~\bibnamefont{Denbleyker}},
  \bibinfo{author}{\bibfnamefont{D.}~\bibnamefont{Du}},
  \bibinfo{author}{\bibfnamefont{Y.}~\bibnamefont{Liu}},
  \bibinfo{author}{\bibfnamefont{Y.}~\bibnamefont{Meurice}}, \bibnamefont{and}
  \bibinfo{author}{\bibfnamefont{A.}~\bibnamefont{Velytsky}},
  \bibinfo{journal}{Phys. Rev.} \textbf{\bibinfo{volume}{D78}},
  \bibinfo{pages}{054503} (\bibinfo{year}{2008}).%, \eprint{0807.0185}.

\bibitem[{\citenamefont{Alves et~al.}(1990{\natexlab{a}})\citenamefont{Alves,
  Berg, and Villanova}}]{alves89}
\bibinfo{author}{\bibfnamefont{N.~A.} \bibnamefont{Alves}},
  \bibinfo{author}{\bibfnamefont{B.~A.} \bibnamefont{Berg}}, \bibnamefont{and}
  \bibinfo{author}{\bibfnamefont{R.}~\bibnamefont{Villanova}},
  \bibinfo{journal}{Phys. Rev.} \textbf{\bibinfo{volume}{B41}},
  \bibinfo{pages}{383} (\bibinfo{year}{1990}{\natexlab{a}}); 
\bibinfo{author}{\bibfnamefont{N.~A.} \bibnamefont{Alves}},
  \bibinfo{author}{\bibfnamefont{B.~A.} \bibnamefont{Berg}}, \bibnamefont{and}
  \bibinfo{author}{\bibfnamefont{S.}~\bibnamefont{Sanielevici}},
  \bibinfo{journal}{Nucl. Phys.} \textbf{\bibinfo{volume}{B376}},
  \bibinfo{pages}{218} (\bibinfo{year}{1992}).%, \eprint{hep-lat/9107002}.

\bibitem[{\citenamefont{Guagnelli et~al.}(1998)\citenamefont{Guagnelli, Sommer,
  and Wittig}}]{guagnelli98}
\bibinfo{author}{\bibfnamefont{M.}~\bibnamefont{Guagnelli}},
  \bibinfo{author}{\bibfnamefont{R.}~\bibnamefont{Sommer}}, \bibnamefont{and}
  \bibinfo{author}{\bibfnamefont{H.}~\bibnamefont{Wittig}}
  (\bibinfo{collaboration}{ALPHA}), \bibinfo{journal}{Nucl. Phys.}
  \textbf{\bibinfo{volume}{B535}}, \bibinfo{pages}{389} (\bibinfo{year}{1998}); 
\bibinfo{author}{\bibfnamefont{S.}~\bibnamefont{Necco}} \bibnamefont{and}
  \bibinfo{author}{\bibfnamefont{R.}~\bibnamefont{Sommer}},
  \bibinfo{journal}{Nucl. Phys.} \textbf{\bibinfo{volume}{B622}},
  \bibinfo{pages}{328} (\bibinfo{year}{2002}).%, \eprint{hep-lat/0108008}.

\bibitem[{\citenamefont{Berlin and Kac}(1952)}]{PhysRev.86.821}
\bibinfo{author}{\bibfnamefont{T.~H.} \bibnamefont{Berlin}} \bibnamefont{and}
  \bibinfo{author}{\bibfnamefont{M.}~\bibnamefont{Kac}},
  \bibinfo{journal}{Phys. Rev.} \textbf{\bibinfo{volume}{86}},
  \bibinfo{pages}{821} (\bibinfo{year}{1952}).

\bibitem[{\citenamefont{Meurice and Zou}()}]{onprogress}
\bibinfo{author}{\bibfnamefont{Y.}~\bibnamefont{Meurice}} \bibnamefont{and}
  \bibinfo{author}{\bibfnamefont{H.}~\bibnamefont{Zou}},
  \emph{\bibinfo{title}{Complex RG flows for 2D nonlinear O(N) sigma models}},
  \bibinfo{note}{preprint in preparation}.

\bibitem[{\citenamefont{Hirsch and Shenker}(1983)}]{PhysRevB.27.1736}
\bibinfo{author}{\bibfnamefont{J.~E.} \bibnamefont{Hirsch}} \bibnamefont{and}
  \bibinfo{author}{\bibfnamefont{S.~H.} \bibnamefont{Shenker}},
  \bibinfo{journal}{Phys. Rev. B} \textbf{\bibinfo{volume}{27}},
  \bibinfo{pages}{1736} (\bibinfo{year}{1983}).

\bibitem[{\citenamefont{Hasenfratz
  et~al.}(1984{\natexlab{a}})\citenamefont{Hasenfratz, Hasenfratz, Heller, and
  Karsch}}]{Hasenfratz:1984hx}
\bibinfo{author}{\bibfnamefont{A.}~\bibnamefont{Hasenfratz}},
  \bibinfo{author}{\bibfnamefont{P.}~\bibnamefont{Hasenfratz}},
  \bibinfo{author}{\bibfnamefont{U.~M.} \bibnamefont{Heller}},
  \bibnamefont{and} \bibinfo{author}{\bibfnamefont{F.}~\bibnamefont{Karsch}},
  \bibinfo{journal}{Phys. Lett.} \textbf{\bibinfo{volume}{B140}},
  \bibinfo{pages}{76} (\bibinfo{year}{1984}{\natexlab{a}}).

\bibitem[{\citenamefont{Yuzhi and Meurice}()}]{hmprogress}
\bibinfo{author}{\bibfnamefont{Y.}~\bibnamefont{Liu}} \bibnamefont{and}
  \bibinfo{author}{\bibfnamefont{Y.}~\bibnamefont{Meurice}},
  \emph{\bibinfo{title}{Complex RG flows for Dyson's hierarchical model}},
  \bibinfo{note}{preprint in preparation}.

\bibitem[{\citenamefont{Alves et~al.}(1990{\natexlab{b}})\citenamefont{Alves,
  Berg, and Sanielevici}}]{alves90b}
\bibinfo{author}{\bibfnamefont{N.~A.} \bibnamefont{Alves}},
  \bibinfo{author}{\bibfnamefont{B.~A.} \bibnamefont{Berg}}, \bibnamefont{and}
  \bibinfo{author}{\bibfnamefont{S.}~\bibnamefont{Sanielevici}},
  \bibinfo{journal}{Phys. Rev. Lett.} \textbf{\bibinfo{volume}{64}},
  \bibinfo{pages}{3107} (\bibinfo{year}{1990}{\natexlab{b}}).

\bibitem[{\citenamefont{Bazavov et~al.}(2009)\citenamefont{Bazavov, Denbleyker,
  Daping, Meurice, Velytsky, and Haiyuan}}]{Bazavov:2009wz}
\bibinfo{author}{\bibfnamefont{A.}~\bibnamefont{Bazavov}},
  \bibinfo{author}{\bibfnamefont{A.}~\bibnamefont{Denbleyker}},
  \bibinfo{author}{\bibfnamefont{D.}~\bibnamefont{Du}},
  \bibinfo{author}{\bibfnamefont{Y.}~\bibnamefont{Meurice}},
  \bibinfo{author}{\bibfnamefont{A.}~\bibnamefont{Velytsky}}, \bibnamefont{and}
  \bibinfo{author}{\bibfnamefont{H.}~\bibnamefont{Zou}},
  \bibinfo{journal}{PoS} \textbf{\bibinfo{volume}{LAT2009}},
  \bibinfo{pages}{218} (\bibinfo{year}{2009}).% \eprint{0910.5785}.
  

\end{thebibliography}

\end{document}